\documentclass[prb,twocolumn,superscriptaddress,amsmath,amssymb,aps,longbibliography,nofootinbib]{revtex4-2}
\usepackage[colorlinks=true,urlcolor=blue,citecolor=blue,linkcolor=blue]{hyperref}
\usepackage{amsmath,amssymb,amsthm}
\usepackage{caption}
\usepackage{float}
\usepackage{graphicx,subfigure}
\usepackage{hyperref}
\usepackage{tikz}
\usepackage{url}
\usepackage{adjustbox}
\usepackage[capitalise]{cleveref}
\usepackage{newtxmath}
\usepackage{bm}
\usepackage[utf8]{inputenc}
\usepackage{multirow}
\usepackage{listings}
\usepackage[linesnumbered, ruled, vlined]{algorithm2e}

\definecolor{psychedelicpurple}{rgb}{0.87, 0.0, 1.0}

\hypersetup{hypertex=true,colorlinks=true,linkcolor=blue,anchorcolor=blue,citecolor=blue} 

\newcommand{\vphi}{{\boldsymbol{\phi}}}
\newcommand{\vx}{{\mathbf{x}}}
\renewcommand{\vv}{{\mathbf{v}}}
\newcommand{\vy}{{\mathbf{y}}}
\renewcommand{\vr}{{\mathbf{r}}}
\crefname{algocf}{alg.}{algs.}
\Crefname{algocf}{Algorithm}{Algorithms}

\begin{document}
\title{Dynamic Hologram Generation with Automatic Differentiation}

\author{Xing-Yu Zhang}
\affiliation{
Hong Kong University of Science and Technology (Guangzhou), Guangzhou 511453, China
}
\affiliation{
Institute of Physics, Chinese Academy of Sciences, Beijing 100190, China
}
\affiliation{
Department of Physics and Astronomy, Ghent University, Krijgslaan 281, 9000 Gent, Belgium
}
\author{Yuqing Wang}
\footnotetext{Xing-Yu Zhang and Yuqing Wang contributed equally to this work.}
\affiliation{
Department of Physics and State Key Laboratory of Low Dimensional Quantum Physics, Tsinghua University, 100084, Beijing, China.
}
\author{Angrui Du}
\affiliation{
Department of Physics and State Key Laboratory of Low Dimensional Quantum Physics, Tsinghua University, 100084, Beijing, China.
}
\author{Han Wang}
\affiliation{
Hong Kong University of Science and Technology (Guangzhou), Guangzhou 511453, China
}
\author{Lei Wang}
\affiliation{
Institute of Physics, Chinese Academy of Sciences, Beijing 100190, China
}
\author{Jinguo Liu}
\email{jinguoliu@hkust-gz.edu.cn}
\affiliation{
Hong Kong University of Science and Technology (Guangzhou), Guangzhou 511453, China
}
\begin{abstract}
    We designed an automatic differentiation-based strategy to generate optical trap arrays that change smoothly in time.
    Instead of repeatedly regenerating the holograms for each time step, we derive the differential form of the phase dynamics that enables the continuous evolution of the trap coordinates.
    This differential form is derived from the implicit differentiation of the fixed point of the Gerchberg-Saxton algorithm, which is computationally efficient.
    We carried out numerical and laboratory experiments to demonstrate its effectiveness in improving the phase continuity and reducing the computational burden compared to the traditional pure interpolation techniques.
    By combining the method with the spatial light modulator, the method is promising for the dynamic manipulation of particles in real experiments.
\end{abstract}
\maketitle

\section{Introduction}
Optical traps~\cite{Ashkin:86, Porter:18} have found a wide range of applications in physics, biophotonics, and biomedicine, including the manipulation of biological cells~\cite{Fazal2011}, the investigation of molecular motor properties~\cite{Malik_2022} and the control of ultracold atoms~\cite{PhysRevX.9.011057}.
The construction and independent manipulation of scalable arrays of uniform optical traps
have become a topic of interest in many fields~\cite{Pang:13, Bahlmann:07, 10.1063/1.1992668, Obata:10, doi:10.1126/science.aah3752, Saffman_2016, Bernien2017}. 
Among the various methodologies proposed to create such arrays~\cite{Pang:13, Bahlmann:07, Gauthier:16}, the computer generated holograms
can be easily implemented with a phase modulated spatial light modulator (SLM)~\cite{Matsumoto:12, Lou2013} which allows
great flexibility in dynamic control~\cite{Hossack:03} and the generation of various geometries~\cite{PhysRevX.4.021034} under certain algorithms~\cite{Gerchberg1972, Chen:17}.

The hologram computation, as illustrated in \Cref{fig: main_figure}, is a process to compute the phase pattern $\vphi$ on the SLM plane that creates traps at coordinates $\vr$.
The efficient and smooth transition of the trap coordinates is crucial for certain applications, such as arranging neutral atoms into a regular pattern~\cite{Wang2023}.
Bottlenecked by the computational cost of the hologram generation algorithm,
the recomputation of the holograms at each step may take too long and cause existing atoms to escape.
Recently, neural network~\cite{Lin2024} and interpolation based methods~\cite{Knottnerus2025} are proposed to improve efficiency.
In this work, we propose a method with better explainability and better numerical stability. Instead of treating the phase-coordinate relationship as a black box, we treat them as an analytic function $\vphi \to \vr$ where $\vphi$ is a matrix of phase in the SLM plane and $\vr$ is a vector of trap coordinates in the image plane.
They are related through Fourier optics~\cite{Goodman2005}. Given a velocity field of the trap locations $\mathbf{v} = \frac{\partial \vr}{\partial t}$, we can derive accurate phase dynamics $\frac{\partial \vphi}{\partial t}$ with automatic differentiation~\cite{Baydin2018}.
Technically, we use the weighted Gerchberg-Saxton (WGS) algorithm~\cite{Gerchberg1972,Kuzmenko2008} to compute the holograms, which iterates (inverse) Fourier transformations between these planes to derive the phases.
This iterative process determines the relation between $\vphi$ and $\vr$ implicitly.
Hence, we employ the implicit function theorem~\cite{krantz2002implicit, Blondel2022} to derive the phase dynamics $\frac{\partial \vphi}{\partial t}$ concerning $\mathbf{v}$.
We emphasize that the process to derive the gradient is computationally efficient, sometimes even faster than the forward computation of the WGS algorithm.
Using the phase dynamics, we can continuously evolve the phases for multiple steps to achieve the contiguous evolution of trap locations.

This paper is organized as follows. In~\cref{subsec: GS algorithm with contiguous Fourier transformation}, we incorporate the contiguous Fourier transformation (CFT) into the WGS algorithm to generate the holograms in the SLM plane at arbitrary trap locations.
In~\cref{Gradient-based evolution}, we derive the phase dynamics $\frac{\partial \vphi}{\partial t}$ concerning trap velocities using the implicit function theorem~\cite{krantz2002implicit, Blondel2022} to differentiate the fixed-point iterations of the WGS algorithm.
In~\cref{sec: Results}, we demonstrate the effectiveness of the proposed method through numerical experiments.
Finally, we implement the proposed setup in a lab experiment and report the results in~\cref{sec:lab}.

To streamline practical implementation, we have developed a Julia package~\cite{slm-flow} that offers pre-built functionality for the evolution of SLM holograms.

\section{Method}

\begin{figure}
    \centering
    \includegraphics[width=0.5\textwidth]{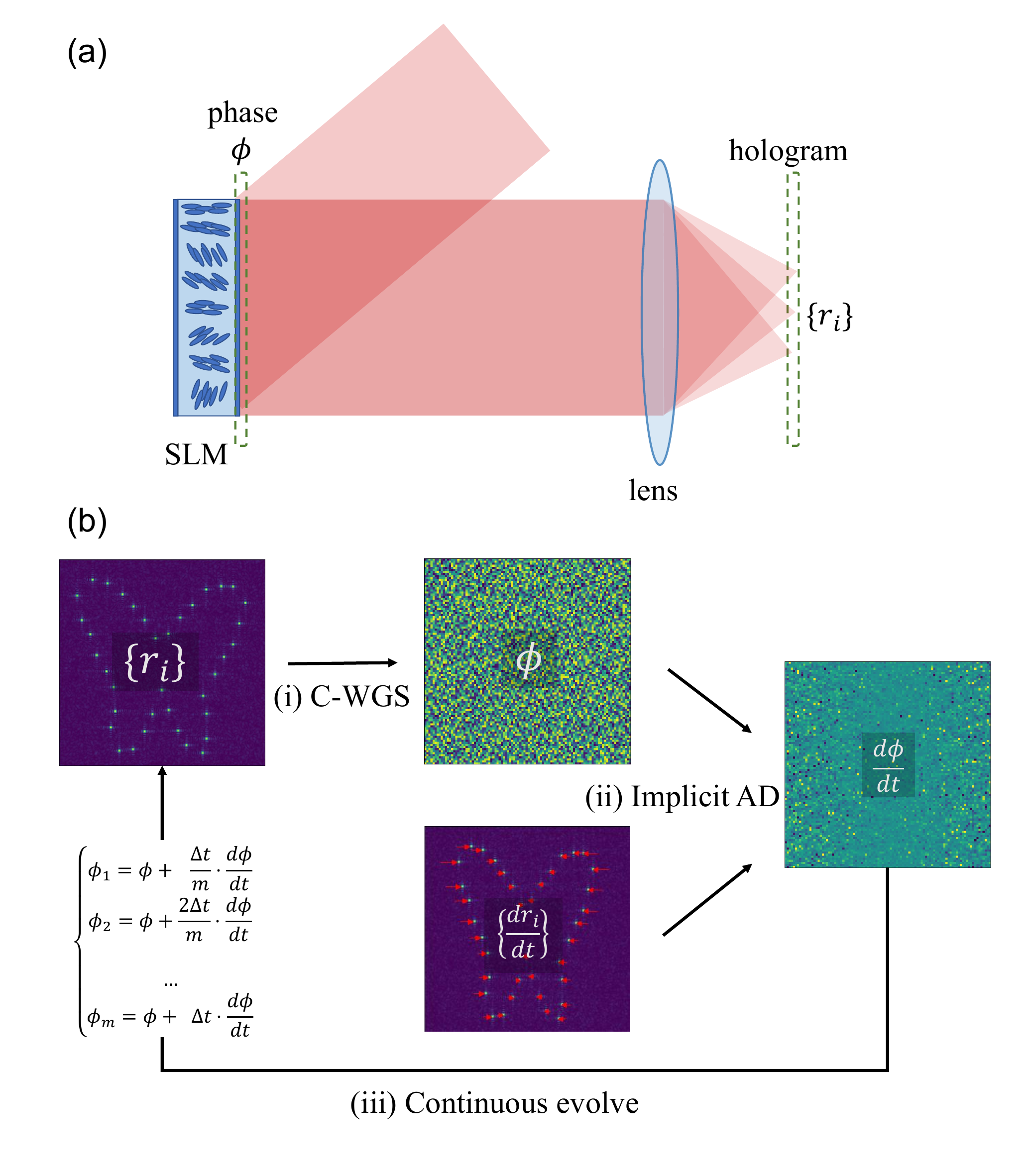}
    \caption{(a) The optical setup for hologram generation. The incident light is phase-modulated by the SLM, which produces the designed hologram at the Fourier plane of the lens. (b) The framework of gradient-flow weighted Gerchberg-Saxton (WGS) algorithm. $\{r_i\}$ is the target trap locations in the focal plane and $\vphi$ is the phase in the SLM plane.
    (i) Use the WGS with contiguous Fourier transformation to generate the phases on the SLM plane. (ii) Given the target velocities of the target trap locations, obtain the phase dynamics $\frac{d\phi}{d t}$ by using the implicit differentiation of the fixed point. (iii) Evolve the phases for multiple steps for contiguous evolution of trap locations.
    }
    \label{fig: main_figure}
\end{figure}

The goal of dynamic hologram generation is to smoothly transition the traps from initial locations to target locations in the image plane. In this work, we intend to achieve this goal with a computer-generated hologram displayed on SLM. When a laser beam is incident on a certain hologram, light gets diffracted. The wavefront or phase profile of the diffracted beam can be modulated by revising the hologram. The framework of our algorithm for calculating the hologram displayed is shown in~\cref{fig: main_figure}.
In our algorithm, this process is divided into $n$ steps, each step evolves the trap locations from one keyframe to the next keyframe, where the keyframe is the trap locations at the $j$-th step denoted as $\vr^{(j)}$, the $i$-th element $r_i^{(j)} \in \mathbb{R}^2$ is a 2D coordinate, and $k$ is the number of traps.
As shown in~\cref{fig: main_figure}, at the beginning of the $j$-th step, we are given the $(j-1)$ -th keyframe in the image plane $\vr^{(j-1)}$ and the target velocity of the $i$-th trap is defined as $\mathbf{v}^{(j)}_i = \frac{r_i^{(j)} - r_i^{(j-1)}}{\Delta t}$ of these trap locations, where $\Delta t$ is the time interval between two steps.
We start by computing the phase generating $\vr^{(j-1)}$ in the SLM plane as $\vphi^{(j-1)} \in \mathbb{R}^{K\times K}$
using the contiguous-WGS (C-WGS) algorithm, where $K$ is the resolution of the SLM plane.
Given the target velocities of the trap locations, we then calculate the phase dynamics $\frac{\partial\vphi^{(j-1)}}{\partial t}$
by automatic differentiation. Finally, we use $\frac{\partial \vphi^{(j-1)}}{\partial t}$ to evolve the phases for multiple steps to achieve the contiguous evolution of trap locations.

In the following subsections, we will introduce how to generate high-resolution holograms using the C-WGS algorithm and how to compute the phase dynamics using automatic differentiation.

\subsection{The Contiguous-WGS algorithm}
\label{subsec: GS algorithm with contiguous Fourier transformation}
The WGS algorithm~\cite{Di2007, Kuzmenko2008,Kim2019a} considers the problem of finding the phase $\vphi$ in the SLM plane that generates the target trap locations $\vr$ in the image plane.
The contiguous WGS (C-WGS) algorithm is a slight modification of the traditional weighted-GS algorithm that incorporates the contiguous Fourier transformation (CFT) to generate high-resolution holograms.

\begin{algorithm}[!ht]
    \caption{The C-WGS algorithm (\texttt{cwgs})}
    \small
    \SetAlgoNoLine
    \SetKwFunction{KwFn}{cwgs\_step}
    \SetKwFunction{KwFnTop}{cwgs}
    \SetKwProg{Fn}{function}{}{end}
    \KwIn{\begin{itemize}
        \item $\mathbf{A_0} \in \mathbb{R}^{K\times K}$: the amplitude in the SLM plane
        \item $\mathbf x \in \mathbb{R}^{k}$ and $\mathbf y \in \mathbb{R}^{k}$: the $x$ and $y$ coordinates of the optical traps
        \item $n \in \mathbb{N}^+$: the total number of iterations\end{itemize}}
    \KwOut{\begin{itemize}
        \item $\vphi \in \mathbb{R}^{K\times K}$: the phase in the SLM plane
        \item $W \in \mathbb{R}^{k}$: the weights applied on the traps.
        \end{itemize}}
    \Fn{\KwFnTop{$\mathbf A_0$, $\mathbf{x}$, $\mathbf{y}$, $n$}}{
        $\vphi_{xy} \sim \mathrm{Uniform}(0, 2\pi)$\tcp*{Random initialize}
        $\mathbf{W}_{u} \leftarrow 1$\tcp*{Initialize trap weights}
        $\mathbf{X}_{ux} \leftarrow e^{-2\pi i u x}$\tcp*{Fourier matrix for $x$-axis}
        $\mathbf{Y}_{vy} \leftarrow e^{-2\pi i v y}$\tcp*{Fourier matrix for $y$-axis}
        \For{j=1,2,\ldots,n}{
            $(\vphi, \mathbf{W})$ $\leftarrow$ \KwFn{$\vphi$, $\mathbf{W}$, $\mathbf{X}$, $\mathbf{Y}$}\;
        }
        \Return $\vphi, \mathbf{W}$\;
    }

    \Fn{\KwFn{$\vphi$, $\mathbf W$, $\mathbf{X}$, $\mathbf{Y}$}}{
        $\mathbf{A}^\prime_{xy} \leftarrow (\mathbf{A}_0)_{xy} e^{i\phi_{xy}}$\tcp*{Update phase}
        $\mathbf{B}_{u} \leftarrow (\mathbf{X}\mathbf{A}^\prime \mathbf{Y}^T)_{uu} $\tcp*{Fourier transform (FT)}
        $\mathbf{W}_{u} \leftarrow \mathbf{W}_{u} \frac{\mathrm{mean}(|\mathbf{B}_{u}|)}{|\mathbf{B}_{u}|}$\tcp*{Update weights}
        $\mathbf{B}^\prime_{u} \leftarrow W_{u}e^{i\mathrm{Arg}(\mathbf{B}_{u})}$\tcp*{Update amplitude}
        $\mathbf{A}_{xy} \leftarrow (\mathbf{X}^\dagger \mathrm{diag}(\mathbf{B}^\prime) \mathbf{Y}^*)_{xy}$\tcp*{Inverse FT}
        $\vphi_{xy} \leftarrow \mathrm{Arg}(A_{xy})$\tcp*{Extract phase}
        \Return $\vphi, \mathbf W$\;
    }
    \label{alg:cwgs} 
\end{algorithm}

The algorithm of contiguous WGS is shown in~\cref{alg:cwgs}. The amplitude in the SLM plane $A_0$ is determined by the light source and cannot be changed during the hologram computation.
Given a set of target trap coordinates $\vr$ in the image plane, the algorithm returns the phase $\vphi$ in the SLM plane.
To ensure that the traps are uniform, the algorithm introduces weights $\mathbf W$ on the traps to adjust the target amplitude of the traps.
The main loop is the same as the traditional WGS algorithm, which is an iterative process (line $7$). We expect the phase $\vphi$ to converge to the fixed point of the \texttt{cwgs\_step} function after $n$ iterations, where the function \texttt{cwgs\_step} defines a self-consistent relationship for $(\vphi, \mathbf W)$.
At the beginning of the function body of \texttt{cwgs\_step}, the wave function on the SLM plane is computed by combining the amplitude $\mathbf{A_0}$ and the phase $\vphi$.
Through Fourier transformation, the algorithm calculates the wave function $\mathbf{B}_u$ at the trap coordinates. The Fourier transformation utilizes the precomputed Fourier matrices $\mathbf X$ and $\mathbf Y$ for the coordinates $x$ and $y$ of the trap locations, respectively.
Then the Fourier transformation can be calculated through matrix multiplication (line $13$ and $16$).
The generated trap amplitudes $\mathbf{B}_u$ may be non-uniform, so the weights $\mathbf{W}_u$ are updated based on the mean of the wave function $\mathbf{B}$ and the current wave function $\mathbf{B}_u$ (line $14$).
The amplitude $\mathbf{B}_u$ is then updated by the weights $\mathbf{W}_u$. The inverse Fourier transformation is then applied to obtain the wave function $\mathbf{A}$ in the SLM plane (line $16$).
Finally, the phase $\vphi$ is extracted from the wave function $\mathbf{A}$ (line $17$).
Instead of using the FFT method, the algorithm calculates the Fourier transformation and its inverse by the matrix multiplication, which has a time complexity of $O(K^2 n_{u})$.
Since in practice $k \ll K \times K$, the computational cost is comparable to the FFT method with complexity $O(K^2 \log(K))$. However, this method does not have the precision issue and zero padding overhead of the FFT method.

As a remark, one can easily generalize the C-WGS algorithm to generate a grid of traps. The computational cost is the same, since the size of the Fourier matrices $\mathbf{X}$ and $\mathbf{Y}$ depends only on the number of unique $\vx$ and $\vy$ coordinates of the traps. Detailed discussion of the grid layout is beyond the scope of this paper.

\subsection{Gradient-based evolution}
\label{Gradient-based evolution}
To temporally evolve the phases $\vphi$ on the SLM plane, 
it is imperative to establish the relationship between the 
phase changing rate $\frac{\partial\vphi}{\partial t} = \frac{\partial \vphi}{\partial \vr} \frac{\partial \vr}{\partial t}$.
Since the trap velocities $\frac{\partial\vr}{\partial t} = \mathbf{v}$ are given, the main task is to compute the Jacobian $\frac{\partial \vphi}{\partial \vr}$.
This relationship can be accurately determined through automatic differentiation (AD)~\cite{Baydin2018} of the C-WGS algorithm in \Cref{alg:cwgs}.
Given sufficient long iterations, the phase in the SLM plane and the target amplitude converge to fixed points $(\vphi^*, \mathbf{W}^*)$ that satisfy the following equation:
\begin{equation}
    (\vphi^*, \mathbf{W}^*) - \texttt{cwgs\_step}(\vphi^*, \mathbf{W}^*, \mathbf{X}, \mathbf{Y}) = \mathbf{0},
\end{equation}
where $\mathbf{X}$ and $\mathbf{Y}$ depends on the trap coordinates $\vr$.
$\vphi^*$ is an implicit function of the trap coordinates $\vr$, the gradient of which can be computed by implicit function theorem~\cite{doi:10.1137/1.9780898717761, krantz2002implicit, Blondel2022}.
For simplicity, we denote $(\mathbf{X}, \mathbf{Y})$ as parameters $\theta$, $(\vphi^*, \mathbf{W}^*)$ as an implicit function $\eta(\theta)$, and $\texttt{cwgs\_step}$ as a function $\mathbf{T}$.
Then the gradient can be computed as:
\begin{equation}\label{eq:series}
    \frac{\partial \eta^*}{\partial \theta} = \sum_{i=0}^{\infty} \left(\frac{\partial \mathbf{T}(\eta^*, \theta)}{\partial \eta^*} \right) ^ {i} \frac{\partial \mathbf{T}(\eta^*, \theta)}{\partial \theta}.
\end{equation}
The derivation of this equation can be found in \Cref{app:implicit}. Intuitively, any fixed point can be viewed as the result of iterating a step infinitely many times. Due to the chain rule, each iteration contributes to the gradient, and the total gradient is the sum of these contributions—this is precisely what the series expansion represents in \cref{eq:series}.
In practice, we can truncate the series in a certain order to obtain an approximate gradient for stability and efficiency.
This Jacobian matrix does not need to be computed explicitly, instead, we use the following relation to compute the disired $\frac{\partial \vphi}{\partial t}$ directly:
\begin{equation}\label{eq:phase-evolution}
    \frac{\partial \vphi}{\partial t} = \left[\frac{\partial \eta^*}{\partial \theta}\right]_{\ \vphi^*} \frac{\partial \theta}{\partial r}  \frac{\partial \vr}{\partial t}.
\end{equation}
Since $\eta(\theta)$ is composed of $\vphi$ and $\mathbf{W}$, we use $\left[\frac{\partial \eta^*}{\partial \theta}\right]_{\ \vphi^*}$ to label the gradient associated with $\vphi^*$.
The above equation can be computed straight-forwardly using the trick in~\Cref{alg:gbe} by utilizing the forward mode AD.
In the algorithm, we introduce a dummy variable $\delta t$ in line 4-5 to represent the infinitesimal time interval. It perturbs the coordinates and this perturbation is reflected in the DFT matrices $\mathbf{X}$ and $\mathbf{Y}$ at lines 6-7. Then the program runs the \texttt{cwgs\_step} function for $m$ times. Since $\vphi$ and $\mathbf{W}$ are already fixed points, they are not changed during the iteration (lines 8-10).
This seemingly trivial computation becomes powerful when combined with the forward mode AD (line 13).
We set the dummy variable $\delta t$ to $0$. It has zero effect to the output, but carries gradient. The AD engine associates each variable with a gradient field, and updates the gradients as the computation goes.
Along with the output, the gradient information $\frac{\partial \vphi}{\partial t}$ is also obtained.
Forward mode AD is efficient in obtaining the gradient of multiple output variables with respect to single input variables, which is suitable for our problem.
AD engines, such as ForwardDiff.jl~\cite{Revels2016} in Julia and PyTorch~\cite{Paszke2019} in Python, can perform forward mode AD and obtain accurate gradients, while only introduce a constant overhead. 
One can verify that the gradient corresponds to~\Cref{eq:phase-evolution} and ~\Cref{eq:series} with the series expansion truncated to the $m$-th order. The finite order error is analyzed numerically as shown in~\cref{fig: more_points_Delta_phi_implicit_expand,fig: more_points_error_aditers}.

\begin{algorithm}[!ht]
    \caption{The gradient-flow algorithm}
    \small
    \SetAlgoNoLine
    \SetKwFunction{KwFna}{gradient\_flow}
    \SetKwFunction{KwFnb}{cwgs\_step}
    \SetKwFunction{KwFnc}{cwgs}
    \SetKwFunction{KwFnf}{cwgs\_iter}
    \SetKwFunction{KwFnF}{ForwardDiff}
    \SetKwProg{Fn}{function}{}{end}
    \KwIn{\begin{itemize}
        \item $\mathbf{A_0} \in \mathbb{R}^{K\times K}$: the amplitude in the SLM plane
        \item $\vx \in \mathbb{R}^{k}$ and $\vy \in \mathbb{R}^{k}$: the $x$ and $y$ coordinates of the optical traps
        \item $n \in \mathbb{N}^+$: the number of iterations to reach fixed point
        \item $m \in \mathbb{N}^+$: the number of series expand in \Cref{eq:series}
    \end{itemize}}
    \KwOut{\begin{itemize}
        \item $\frac{d\vphi}{dt} \in \mathbb{R}^{K\times K}$: the change rate of phase
        \end{itemize}}
    \Fn{\KwFna{$\mathbf{A}_0$, $\vx$, $\vy$, $\vv_x$, $\vv_y$, $n$, $m$}}{
        ($\vphi$, $\mathbf{W}$) $\leftarrow$ \KwFnc{$\mathbf{A_0}$, $\mathbf{x}$, $\mathbf{y}$, $n$}\tcp*{Get fixed point}
        \Fn{\KwFnf{$\delta t$}}{
        $\vx \leftarrow \vx + \vv_x \delta t$ \tcp*{Infinitesimal movement}
        $\vy \leftarrow \vy + \vv_y \delta t$\;
        $\mathbf{X}_{ux} \leftarrow e^{-2\pi i u x}$\tcp*{DFT matrices}
        $\mathbf{Y}_{vy} \leftarrow e^{-2\pi i v y}$\;
        \For{j=1,2,\ldots,m}{
        $(\vphi, \mathbf{W})$ $\leftarrow$ \KwFnb{$\vphi$, $\mathbf{W}$, $\mathbf{X}$, $\mathbf{Y}$}\;
        }
        \Return $\vphi$\;
        }
        $\frac{d\vphi}{dt} \leftarrow \KwFnF(\KwFnf, \delta t=0)$ \tcp*{Calculate $\frac{d\vphi}{d t}$ using forward mode AD}
        \Return $\frac{d\vphi}{d t}$;
    }
    
    \label{alg:gbe} 
\end{algorithm}

\section{Benchmarks and Applications}
\label{sec: Results}
In this section, we will present specific examples and benchmark our method to demonstrate the improvements.

\subsection{Movement in One Direction}
\begin{figure}
    \centering
    \includegraphics[width=0.5\textwidth]{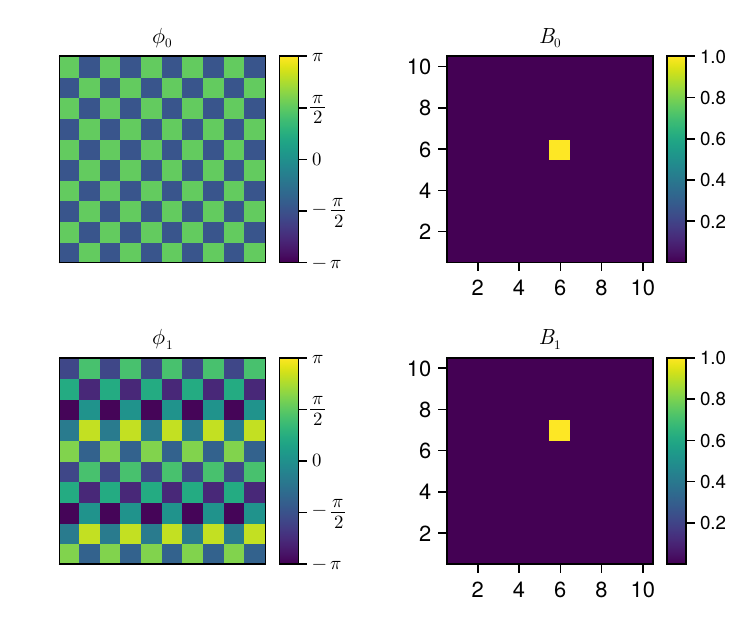}
    \caption{The patterns $B$ in the focal plane and the corresponding phases $\phi$ in the SLM plane of the exact solution. $\phi_0$ and $B_0$ represent the initial state, while $\phi_1$ and $B_1$ depict the final state.}
    \label{fig: one_point_phi_B}
\end{figure}

In this section, we plan to demonstrate the gradient evolution method through a single optical trap moving in one direction. We start with an SLM of size $10 \times 10$, where the initial configuration features a single trap positioned at $(0.5, 0.5)$, which then moves upward by $0.1$ in the $y$ direction, resulting in the final position at $(0.5, 0.6)$. Let us denote the initial and final phases in the SLM plane as $\phi_0$ and $\phi_1$, respectively. These two simple phases can be easily solved, as illustrated in~\cref{fig: one_point_phi_B}.

\begin{figure}
    \centering
    \includegraphics[width=0.5\textwidth]{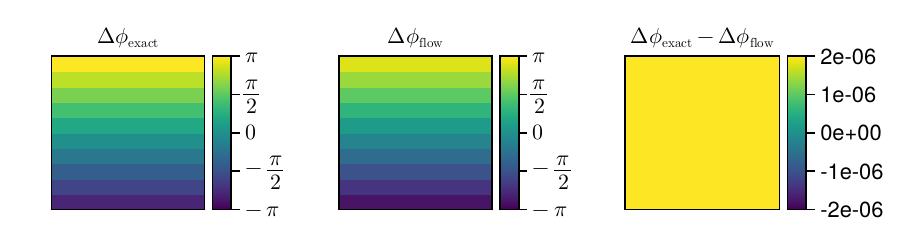}
    \caption{Comparison of the exact phase change with the current gradient flow method.}
    \label{fig: one_point_Delta_phi_implicit}
\end{figure}

We can assess the phase change $\Delta \phi = \phi_1 - \phi_0$ resulting from the current gradient flow method compared to the exact solution. The discrepancy, denoted by $\Delta \phi_{\rm{exact}} - \Delta \phi_{\rm{flow}}$, is at the level of $10^{-6}$, as shown in \cref{fig: one_point_Delta_phi_implicit}.

As shown in the previous example, the displacement distance $\Delta x$ is directly proportional to $\Delta \phi$, enabling the WGS algorithm to attain the exact solution.
However, in situations with more complex patterns, the WGS algorithm can only offer phase approximations within the SLM plane. As a result, discrepancies in phases may arise even with small movements. An example demonstrating this is presented in~\cref{sec: Four-point movement}

In order to decrease the error in the gradient calculation, we expand the series of the derivative derived from the implicit function theorem and choose a higher number of series, which can be demonstrated through~\cref{eq: expand_series}. 
 In practical applications, employing approximately 15 series is sufficient to obtain an accurate gradient, as depicted in~\cref{fig: more_points_Delta_phi_implicit_expand,fig: more_points_error_aditers}.

\subsection{Squeezing along one direction}\label{subsec:squeeze}

In this section, we intend to assess the numerical quality of $\Delta \phi$, which can indicate to what extent we can use $\Delta \phi$ to evolve $\phi$ and follow trap movement.

\begin{figure}
    \centering
    \includegraphics[width=0.5\textwidth]{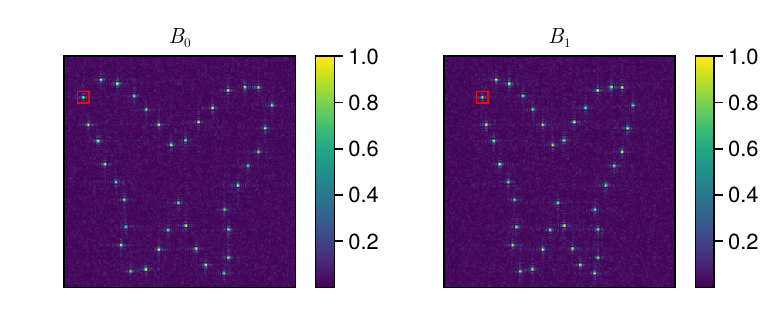}
    \caption{Patterns of the butterfly denoted as $B$ in the focal plane. $B_0$ represents the initial state, whereas $B_1$ signifies the final state. Various points undergo distinct displacements, with the point exhibiting the longest displacement highlighted by a red box.}
    \label{fig: butterfly_B}
\end{figure}

In the corresponding simulation, the initial image is designed to be a butterfly, while the target diagram is the butterfly with shrinking wings, which means that the points on both sides converging towards the center. The simulation process is illustrated in~\cref{fig: butterfly_B}.

At first, we employed a grid size of $100\times 100$ for the SLM and a focal plane with dimensions $1000\times 1000$ for further analysis. We focus on the top-left point marked with a red box, which experiences a substantial displacement of approximately 0.08 units, corresponding to 8 pixels on the present SLM.

\begin{figure}[!ht]
    \centering
    \includegraphics[width=0.5\textwidth]{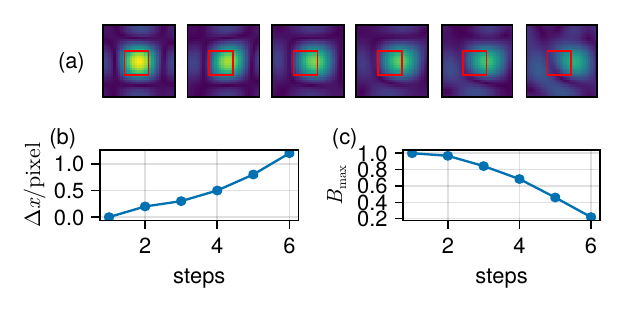}
    \caption{(a) Utilizing the gradient to guide the optical trap results in a gradual decay of the amplitude of the point in~\cref{fig: butterfly_B}. The red box labels one unit pixel. (b)(c) The displacement $\Delta x$ per pixel of the central points and the maximum amplitude $B_{\mathrm{max}}$ over successive steps in (a).}
    \label{fig: butterfly_flow_decay}
\end{figure}

After a single round of WGS calculation, we can derive the gradient $\frac{d \phi}{d t}$ of fixed points according to the second line of~\cref{eq: expand_series} instead of solving the linear equation in the first line for greater stability and efficiency. Then this gradient is applied in the phase evolution process $\phi$ in the SLM plane. However, with increasing time or distance applied to the gradient, the amplitude $B$ of the image in the focal plane will gradually decay, as shown in~\cref{fig: butterfly_flow_decay}(a). To quantitatively analyze the decay phenomena, we plot the maximum amplitude $B_{\mathrm{max}}$ of the points influenced with moving distance $\Delta x$ per pixel of the center along the six steps in~\cref{fig: butterfly_flow_decay}(a). We find that the amplitude decays rapidly from 0.7 to 0.2 after 0.5 to 1 pixel movement (the initial amplitude is 1), which is depicted in~\cref{fig: butterfly_flow_decay} (b) and (c). Therefore, we recommend recalculating the keyframes after a 0.5-pixel movement to prevent significant amplitude decay. The long-distance result is shown in the red-box point in~\cref{sec: Full movement of the butterfly}

\subsection{Moving a subset of traps}
\begin{figure}[!ht]
    \centering
    \includegraphics[width=0.5\textwidth]{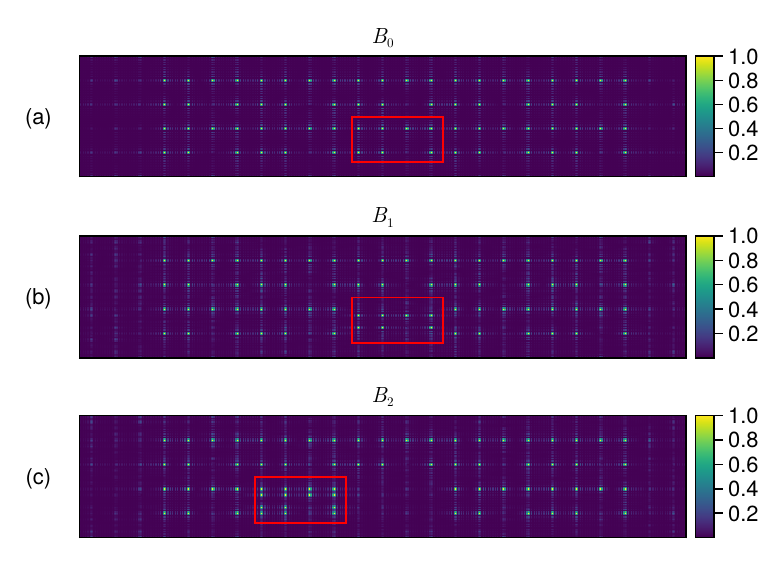}
    \caption{$B_0$ is the initial configuration of the logical quantum processor, while $B_1$ is the shrunk middle configuration and $B_2$ represents the final states after the logical operations.}
    \label{fig: logical_quantum_processor_B}
\end{figure}

\begin{figure}[!ht]
    \centering
    \includegraphics[width=0.5\textwidth]{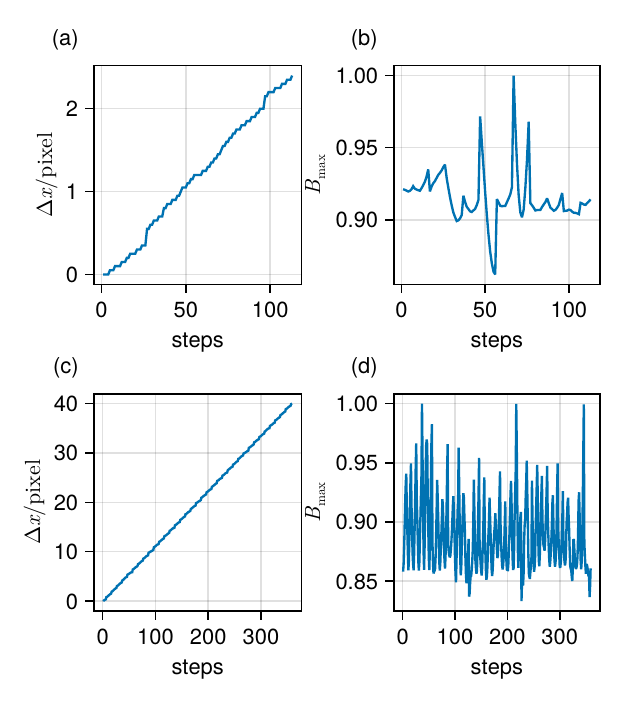}
    \caption{The displacement $\Delta x$ per pixel of the central points and the maximum amplitude $B_{\mathrm{max}}$ over successive steps of (a)(b) $B_0$ to $B_1$ and (c)(d) $B_1$ to $B_2$ in~\cref{fig: logical_quantum_processor_B}. }
    \label{fig: logical_quantum_processor_flow_all}
\end{figure}

The current methodology can be applied in the manipulation of a logical quantum processor~\cite{Bluvstein2024}. Initially, we focus on a cluster of points arranged in a $2\times 5$ unit cell of seven points shown in ~\cref{fig: logical_quantum_processor_B} (a). We performed a logical calculation involving the shrinking of the width along $y$ direction in~\cref{fig: logical_quantum_processor_B} (b) and then moved the unit cell to the left side(shown in~\cref{fig: logical_quantum_processor_B} (c).). The displacement $\Delta x$ per pixel of the central points and the maximum amplitude $B_{\mathrm{max}}$ over successive steps ($B_0,B_1,B_2$) are illustrated in~\cref{fig: logical_quantum_processor_flow_all}. The results demonstrate a new approach for seamless transformation of the logical quantum processor.

\subsection{Rearrangement of traps}
\begin{figure}[!ht]
    \centering
    \includegraphics[width=0.5\textwidth]{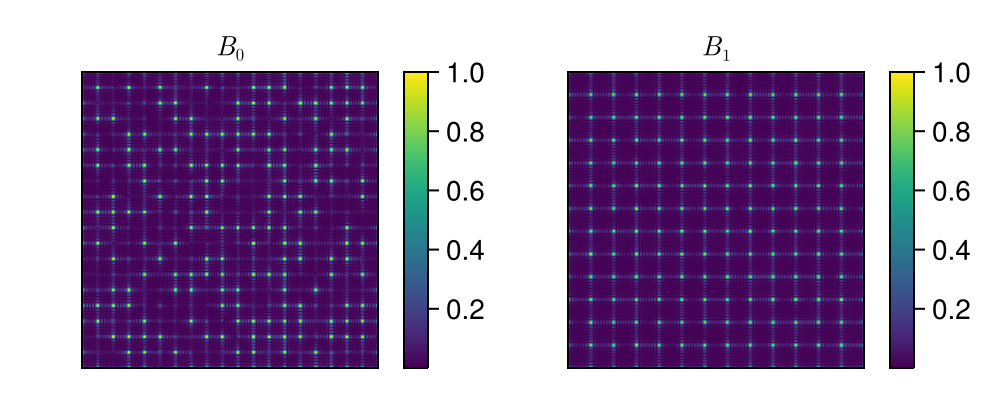}
    \caption{$B_0$ is the initial configuration of the half-filled atoms, while $B_1$ represents the final states of the full-filled atoms. }
    \label{fig: half_fill_to_full_fill}
\end{figure}

\begin{figure}[!ht]
    \centering
    \includegraphics[width=0.5\textwidth]{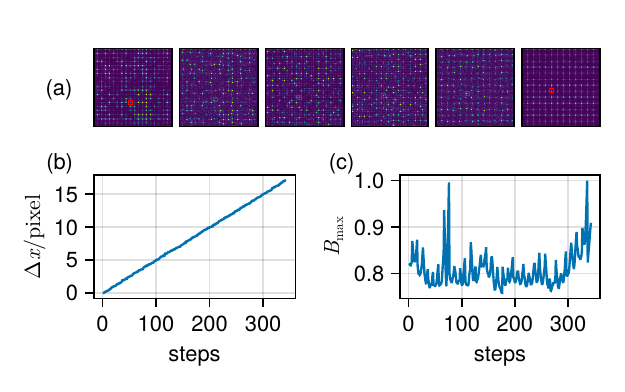}
    \caption{(a) Utilizing the gradient to guide the optical trap results in a gradual decay of the amplitude of the point in~\cref{fig: half_fill_to_full_fill}. (b)(c) The displacement $\Delta x$ per pixel of the red boxed points and the maximum amplitude $B_{\mathrm{max}}$ over successive steps. }
    \label{fig: half_fill_to_full_fill_all}
\end{figure}

Rearranging the position of traps is crucial for atom loading. In this example, we first generate a randomly positioned about half-filled $20\times20$ grid (256 points by randomly picked) to a full-filled $16\times16$ grid in square pattern (shown in~\cref{fig: half_fill_to_full_fill}). We employ the Hungarian algorithm~\cite{kuhn1955hungarian} to match the points in the two configurations, ensuring that the total displacement is minimized. Compared with methods of moving from outside to get a fully filled pattern~\cite{manetsch2024tweezerarray6100highly}, this manipulation is an efficient alternative way to obtain the desired pattern in experiments~\cite{PhysRevApplied.22.024073}. We also calculated the displacement $\Delta x$ per pixel of the central points and the maximum amplitude $B_{\mathrm{max}}$ over successive steps (depicted in~\cref{fig: half_fill_to_full_fill_all}) to demonstrate the feasibility of the current methodology.

\subsection{Performance analysis}\label{sec:benchmark}
Our computational setup involves the Nvidia Ampere A800 GPU, boasting a substantial computing capacity of 9.7 tera floating point operations per second (TFLOPS) and 1.5 terabytes of memory bandwidth per second. Additionally, the CPU utilized is the AMD EPYC 7702 64-Core Processor @ 3.35GHz, offering a single-thread single-precision computing power of 195 giga floating point operations per second (GFLOPS). We use the Julia package ForwardDiff.jl~\cite{RevelsLubinPapamarkou2016} for automatic differentiation and the CUDA.jl package~\cite{besard2018juliagpu} for GPU acceleration.

A benchmark of the computation time for calculating the WGS and gradient of the butterfly and circle transformations in \cref{tab: WGS_AD_benchmark}. It is observed that the computational time of GPU processing is significantly smaller compared to CPU processing, which is attributed to the proficiency of GPUs in handling matrix manipulations. It is a key advantage leveraged by the current contiguous Fourier transformation methodology.
The time to compute gradients can be much less than the WGS computation, which is a promising result for the future development of the methodology.

\begin{table}
    \begin{tabular}{l|l|l|l}
    \hline\hline
    Program & Implementation & butterfly & circle \\ \hline
    \multirow{2}{*}{WGS}         & CPU                        & 3.38                           & 2.87                       \\
                                 & GPU                        & 0.0375                         & 0.0795                     \\ \hline
    \multirow{4}{*}{Gradient}    & CPU Linear Solve           & 7.16                           & 79.4                       \\
                                 & CPU Expand                 & 2.26                           & 25.1                       \\
                                 & GPU Linear Solve           & 0.0268                         & 0.179                      \\
                                 & GPU Expand                 & \textbf{0.00861}     & \textbf{0.0586}  \\ \hline\hline
    \end{tabular}
    \caption{Wall clock time in seconds for hologram computation with WGS algorithm and gradient computation on a $1024 \times 1024$ grid SLM. The WGS iterations are set at 50, and the expansion series is truncated to order 15. All simulations are conducted using the Float64 data type. }
    \label{tab: WGS_AD_benchmark}
\end{table}

\section{Experiment}\label{sec:lab}
To benchmark our method's performance, we experimentally simulated trap rearrangement using a LCOS-SLM (Hamamatsu, X15213-02L). The SLM has a resolution of $1272 \times 1024$ and a frame rate of 60 Hz, limited by the DVI transmission rate. We randomly generated initial trap positions on an 18×18 square grid with a filling factor of 0.44. We then calculated the phase using our method and projected it onto the SLM in real time to rearrange the traps into a defect-free $12 \times 12$ configuration. We accelerated phase generation on an NVIDIA RTX 4090 GPU. Phase generation takes 75 ms per 10 frames, including 20 iterations of WGS calculation and gradient evolution. Projection requires 128 ms per 10 frames, which aligns with the standard frame rate but is not fast enough compared to the phase generation.

\begin{figure}[!ht]
    \centering
    \includegraphics[width=0.5\textwidth]{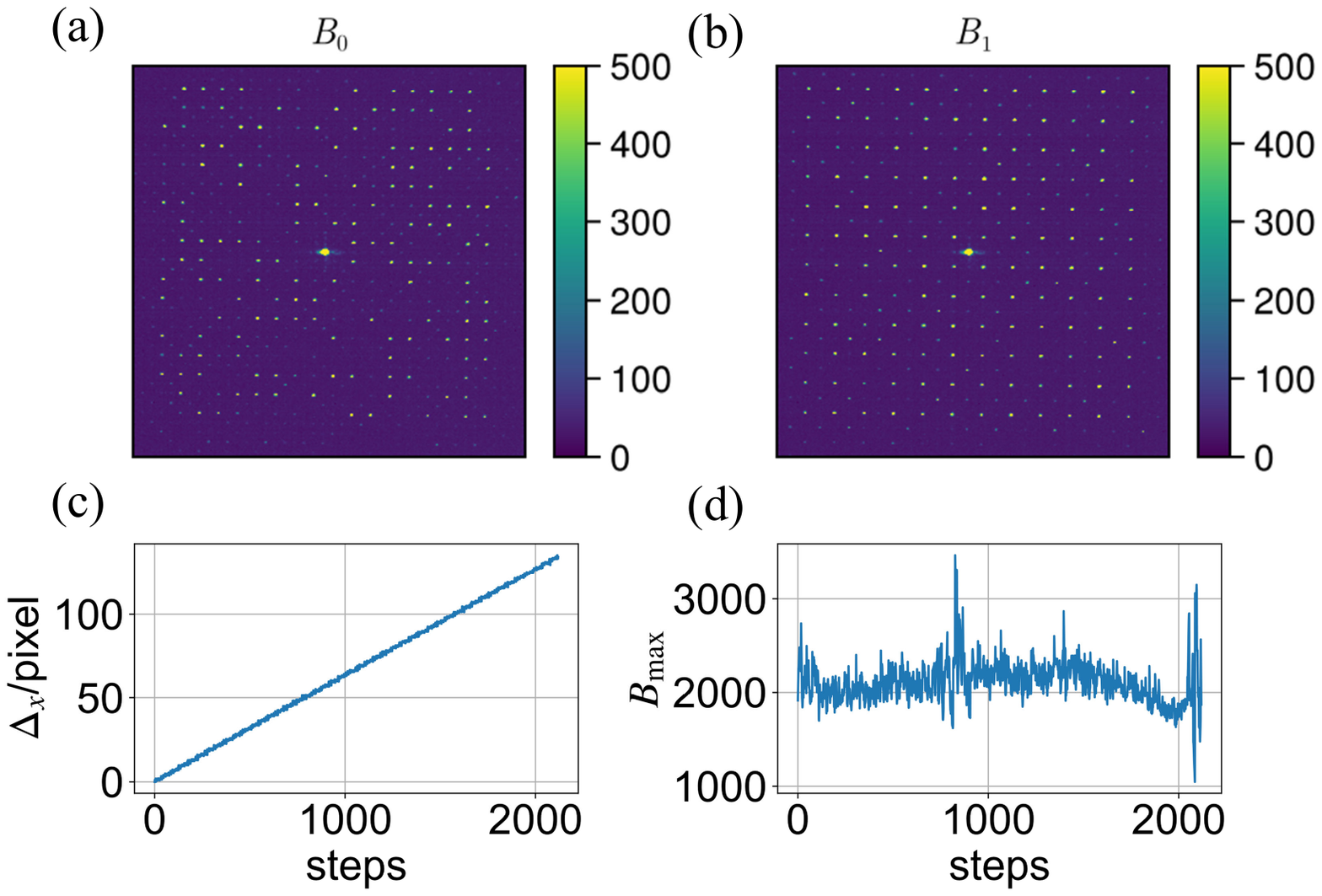}
    \caption{(a) The first image before rearrangement, showing randomly initialized optical traps. The bright spot in the center is the zero-order light that remains unmodulated by the SLM. (b) The defect-free $12\times12$ optical traps after rearrangement. (c)(d) The displacement $\Delta x$ per pixel and the maximum intensity $B_{\mathrm{max}}$ of one optical trap across successive steps. Data was extracted from a monitor camera, with displacement measured in camera pixels and intensity measured in camera gray values.}
    \label{fig: exp_rearrange}
\end{figure}

Throughout each movement, we monitored the positions and intensities of optical tweezers using a CMOS camera in the Fourier plane. \cref{fig: exp_rearrange} shows the displacement $\Delta x$ and maximum intensity $B_{\mathrm{max}}$ of one traced optical trap. Using realistic experimental intensity fluctuations, we simulated the decay of atoms in optical traps similar to~\cite{Lin2024}, more informations see~\cref{sec: atoms_decay}. The results demonstrate that the optical traps can be rearranged without significant intensity decay, indicating the effectiveness of our method for real-time optical trap manipulation.

\section{Discussion}
In summary, integrating automatic differentiation with the weighted Gerchberg-Saxton (WGS) algorithm enables seamless manipulation of optical focus arrays. The numerical analysis demonstrates that, in the case of single-point movement, the gradient can accurately adjust the trap locations by any desired distance.
In other scenarios, the gradient can accurately adjust the image by at least 0.5 pixels without significant amplitude decay, making it suitable for various pattern movements. While 0.5 pixels may not represent a large displacement, it is important to note that the gradient computation is highly efficient, requiring even less time than the hologram computation itself. By leveraging the gradient, we can enhance the continuity of trap coordinate changes, which may help reduce atom loss during the loading process in realistic atom array experiments.
Compared to previous methods, a key advantage of our approach is its explainability, along with its easy extensibility to 3D GS algorithms or any other fixed-point-iteration-based method. The gradient can be easily understood and analyzed, which is crucial for experimentalists to comprehend why the method succeeds or fails.

The effectiveness of the method is also demonstrated through a laboratory experiment. Here, we identify two potential areas for improvement in the devices.
First is the frame rate. In comparison to acousto-optic deflectors, which can move traps on a microsecond timescale, the current setup's frame rate of 60 Hz represents a significant bottleneck, limiting its application in atom loading experiments.
The second area for improvement is the precision of phase adjustment. In practice, the spatial light modulator offers only 8 or 12 bits of precision, which falls short of the continuous limit. Achieving seamless phase manipulation remains a significant challenge for experimentalists. Given that our proposed method is characterized by its explainability, simplicity, and computational efficiency, we hope it will inspire further research aimed at overcoming these limitations.

\section{Acknowledgment}
We acknowledge the very helpful discussion with Wenlan Chen. This project is partially supported by the National Natural Science Foundation of China under grant nos. 92270107, 12404568, T2225018, and T2121001, and the Guangzhou Municipal Science and Technology Project (No. 2024A03J0607)

\appendix

\section{Implicit function theorem}\label{app:implicit}
We consider a user-defined mapping $\mathbf{F}: \mathbb{R}^d \times \mathbb{R}^n \rightarrow \mathbb{R}^d$, and an implicit function $\eta^*(\theta)$ that satisfies the following equation:
\begin{equation}
    \mathbf{F}(\eta^*(\theta), \theta) = 0.
    \label{eq: Ffunction}
\end{equation}
The implicit function theorem states that if a point $(\eta_0, \theta_0)$ satisfies $F(\eta_0, \theta_0) = 0$ with a continuously differentiable function $\mathbf{F}$, and the Jacobian $\frac{\partial\mathbf{F}}{\partial \eta}$ at $(\eta_0, \theta_0)$ is an invertible square matrix, then there exists a function $\eta(\cdot)$ defined in a neighborhood of $\theta_0$ such that $\eta^*(\theta_0) = \eta_0$. Furthermore, for all $\theta$ in this neighborhood, $\mathbf{F}(\eta^*(\theta), \theta) = 0$ and $\frac{\partial \eta^*}{\partial \theta}$ exist. Under the chain rule, the Jacobian $\frac{\partial \eta^*}{\partial \theta}$ satisfies:
\begin{equation}
    \frac{\partial \mathbf{F}(\eta^*, \theta)}{\partial \eta^*} \frac{\partial \eta^*}{\partial \theta} + \frac{\partial \mathbf{F}(\eta^*, \theta)}{\partial \theta} = 0.
\end{equation}
calculating $\frac{\partial \eta^*}{\partial \theta}$ involves solving the system of linear equations expressed as
\begin{equation}
    \underbrace{\frac{\partial \mathbf{F}(\eta^*, \theta)}{\partial \eta^*}}_{V \in\mathbb{R}^{d \times d}} \underbrace{\frac{\partial \eta^*}{\partial \theta}}_{J \in\mathbb{R}^{d \times n}}=-\underbrace{\frac{\partial \mathbf{F}(\eta^*, \theta)}{\partial \theta}}_{P \in\mathbb{R}^{d \times n}}. 
    \label{eq: implicit_linear_eqution}
\end{equation}
Therefore, the desired Jacobian is given by $J=V^{-1}P$.
In many practical situations, explicitly constructing the Jacobian matrix is unnecessary. Instead, it suffices to perform left-multiplication or right-multiplication by $V$ and $P$. These operations are known as the vector-Jacobian product (VJP) and the Jacobian-vector product (JVP), respectively. They are valuable for determining $\eta(\theta)$ using reverse-mode and forward-mode automatic differentiation (AD), respectively.

In many cases, $\mathbf{F}$ is clearly defined, allowing for facilitated VJP or JVP calculations using AD. However, there are instances where $\mathbf{F}$ is implicitly defined, as seen in problems involving variational problem. In such cases, determining the VJP or JVP will require implicit differentiation, a method known as automatic implicit differentiation~\cite{Blondel2022}.
In our scenario, the function $\eta^*(\theta)$ is implicitly defined through a fixed point equation:
\begin{equation}
    \eta^*(\theta) = \mathbf{T}(\eta^*(\theta), \theta),
\end{equation}
where $\mathbf{T}: \mathbb{R}^d \times \mathbb{R}^n \rightarrow \mathbb{R}^d$. This representation can be viewed as a specific instance of~\cref{eq: Ffunction} by introducing the residual term:
\begin{equation}
    \mathbf{F}(\eta, \theta) = \mathbf{T}(\eta, \theta) - \eta.
\end{equation}
If $\mathbf{T}$ is continuously differentiable, application of the chain rule yields:
\begin{align}
    V &= -\frac{\partial \mathbf{F}(\eta^*, \theta)}{\partial \eta^*} = I - \frac{\partial \mathbf{T}(\eta^*, \theta)}{\partial \eta^*}, \label{eq: implicit_linear_eqution_TV}\\
    P &= \frac{\partial \mathbf{F}(\eta^*, \theta)}{\partial \theta} =  \frac{\partial \mathbf{T}(\eta^*, \theta)}{\partial \theta}, \label{eq: implicit_linear_eqution_TP}
\end{align}
Substituting $V$ and $P$ back into~\cref{eq: implicit_linear_eqution}, we obtain:
\begin{equation}
    \begin{split}
     \frac{\partial \eta^*}{\partial \theta} &= \left(I - \frac{\partial \mathbf{T}(\eta^*, \theta)}{\partial \eta^*} \right) ^ {-1} \frac{\partial \mathbf{T}(\eta^*, \theta)}{\partial \theta} 
     \\
     &= \sum_{i=0}^{\infty} \left(\frac{\partial \mathbf{T}(\eta^*, \theta)}{\partial \eta^*} \right) ^ {i} \frac{\partial \mathbf{T}(\eta^*, \theta)}{\partial \theta}
    \end{split}
     \label{eq: expand_series}
\end{equation}
We can use either solve the linear equation in the first line or expand the series in the second line to compute the gradient. The latter is more stable and efficient in practice.
Consequently, when differentiating a fixed-point iteration, the main operation involves the JVP related to the single-step iteration function $\mathbf{T}$. In practical situations, we can iterate $\mathbf{T}$ over multiple steps, obtaining the derivative for each step, leading to an accumulation of the expanded series. Although this method increases memory usage, it still shows effectiveness in real-world applications~\cite{PhysRevX.9.031041, PhysRevB.108.085103}.

To provide further clarification, we employ the WGS algorithm as our fixed-point function $\mathbf{T}((\vphi, \mathbf{W}), \theta)$, which encompasses multiple steps outlined as~\Cref{alg:cwgs} and with the implicit function $\eta(\theta)=(\vphi(\theta), \mathbf{W}(\theta))$ and $\theta(r)=(\mathbf{X}(r), \mathbf{Y}(r))$.

To address the solution for $\frac{\partial\vphi}{\partial t}$, we have the following:
\begin{equation}
\begin{split}
\frac{\partial\vphi}{\partial t}&=\frac{\partial\vphi}{\partial \theta}\frac{\partial\theta}{\partial \vr}\frac{\partial \vr}{\partial t}
=\left[\frac{\partial\eta}{\partial \theta}\right]_{\vphi}\frac{\partial\theta}{\partial \vr}\frac{\partial \vr}{\partial t}\\
&=\left[V^{-1}P\right]_{\vphi}\frac{\partial\theta}{\partial \vr}\frac{\partial \vr}{\partial t}
\end{split}
\end{equation}
where we use $[\cdot]_\vphi$ to label the part associated with $\vphi$ and the corresponding linear coefficients in~\cref{eq: implicit_linear_eqution_TV} and ~\cref{eq: implicit_linear_eqution_TP} should be revised as:
\begin{align}
    V & = I - \frac{\partial \mathbf{T}((\vphi^*, \mathbf{W^*}), \theta)}{\partial (\phi^*, \mathbf{W^*})}, \\
    P &= \frac{\partial \mathbf{T}((\vphi^*,\mathbf{W^*}), \theta)}{\partial \theta}. \label{eq: linear_coefficient_B}
\end{align}

\section{Four points movement}
\label{sec: Four-point movement}
\begin{figure}[!ht]
    \centering
    \includegraphics[width=0.5\textwidth]{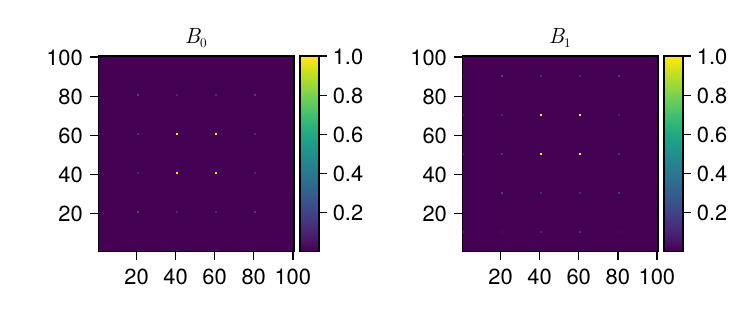}
    \caption{Patterns of the four points denoted as $B$ in the focal plane. $B_0$ represents the initial state, while $B_1$ signifies the final state after a $0.1$ movement in the $y$ direction.}
    \label{fig: more_points_B}
\end{figure}

\begin{figure}[!ht]
    \centering
    \includegraphics[width=0.5\textwidth]{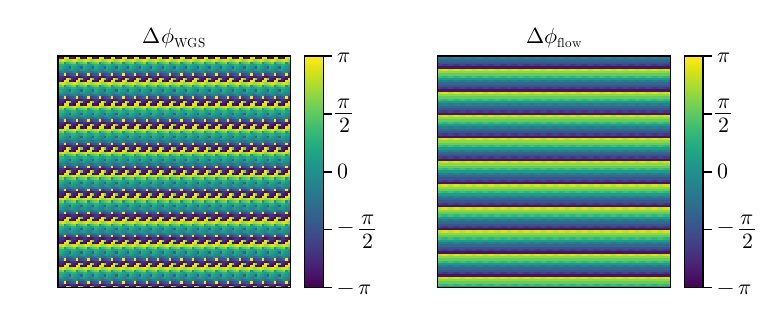}
    \caption{The phase difference $\Delta \phi$ between two states. The $\Delta \phi_{\rm{flow}}$ obtained from the current method exhibits a notably smoother trend compared to the $\Delta \phi_{\rm{WGS}}$ derived from the WGS algorithm.}
    \label{fig: more_points_Delta_phi_wgs_flow}
\end{figure}

Consider the scenario involving four points located at $(0.4, 0.4), (0.4, 0.6), (0.6, 0.4), (0.6, 0.6)$, transitioning to $(0.4, 0.5), (0.4, 0.7), (0.6, 0.5), (0.6, 0.7)$ with a $0.1$ movement in the $y$ direction, as depicted in~\cref{fig: more_points_B} with a SLM grid SLM $100\times100$. When trying to derive the phases directly using the WGS algorithm, the resulting $\Delta \phi$ values exhibit a significantly larger magnitude than those obtained through the current gradient flow method, even when initialized from one another. This discrepancy is visually evident in~\cref{fig: more_points_Delta_phi_wgs_flow}.

\begin{figure}[!ht]
    \centering
    \includegraphics[width=0.5\textwidth]{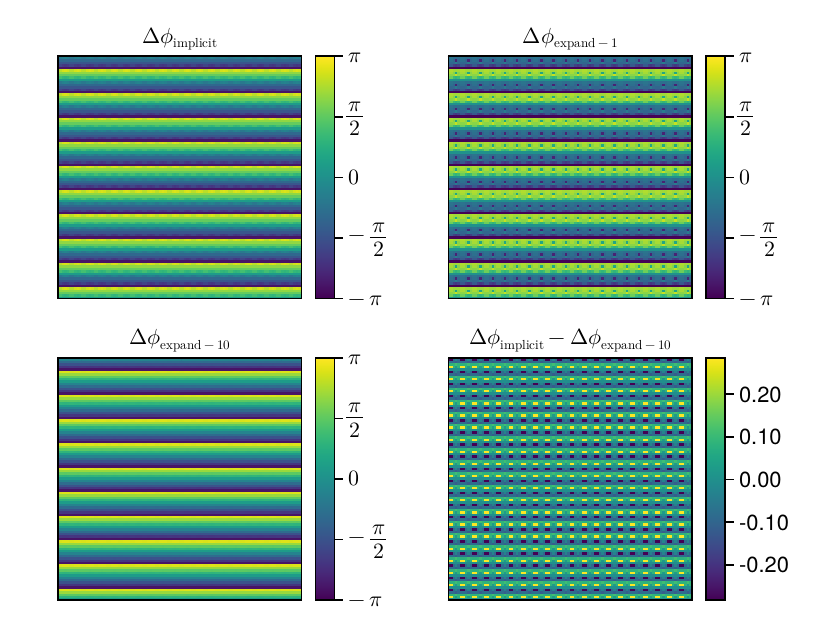}
    \caption{The phase difference $\Delta \phi_{\mathrm{implicit}}$ obtained from the solution of the linear equation and $\Delta \phi_{\mathrm{expand}}$ from the series expansion in~\cref{eq: expand_series}. The label $-n$ denotes the number of series used.}
    \label{fig: more_points_Delta_phi_implicit_expand}
\end{figure}

\begin{figure}[!ht]
    \centering
    \includegraphics[width=0.45\textwidth]{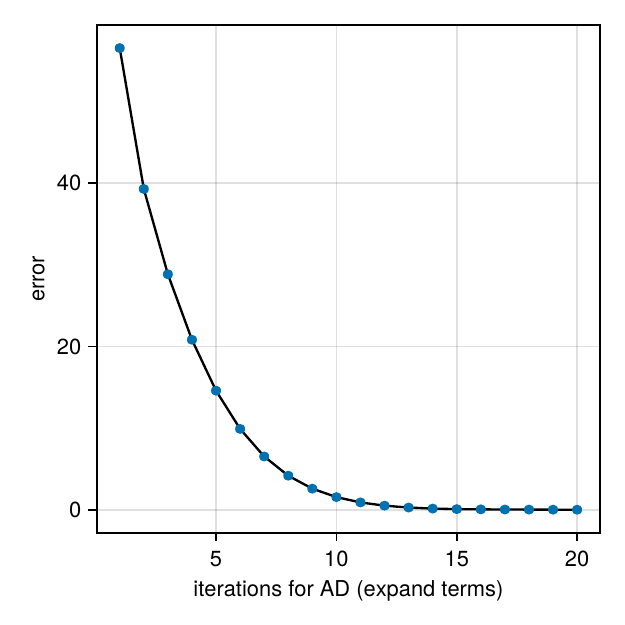}
    \caption{The error across iterations for AD. With an increase in the number of expanded terms, the error significantly diminishes. The error is defined as $|\Delta \phi_{\rm{implicit}} - \Delta \phi_{\rm{expand}}|$.}
    \label{fig: more_points_error_aditers}
\end{figure}

\section{Full movement of the butterfly}
\label{sec: Full movement of the butterfly}
\begin{figure}[!ht]
    \centering
    \includegraphics[width=0.5\textwidth]{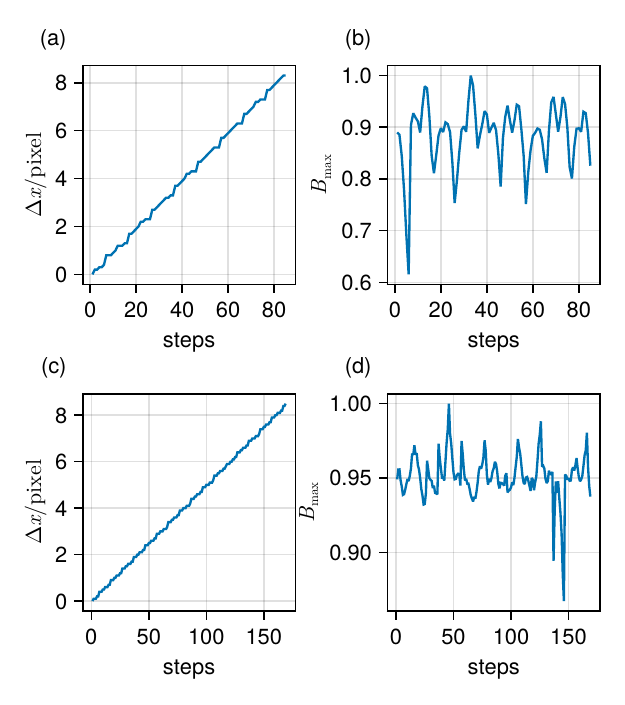}
    \caption{(a)(b) The displacement $\Delta x$ per pixel and the maximum amplitude $B_{\mathrm{max}}$ of the central points over successive steps in the evolution depicted in~\cref{fig: butterfly_B}, involving a 1-pixel recalculation of the WGS keyframes. (c)(d) 0.5-pixel recalculation.}
    \label{fig: butterfly_flow_all}
\end{figure}

Based on the aforementioned analysis in~\cref{subsec:squeeze}, we can recalculate the WGS step following a displacement of 0.5 pixels of the furthest moving point. An effective strategy involves leveraging the gradient to regress the previous 0.5 pixels, thus necessitating recalculation only once per 1-pixel movement interval. Consequently, during the transition from $B_0$ to $B_1$ in~\cref{fig: butterfly_B}, it is advisable to recalculate the 8th WGS algorithm, incorporating 10 gradient flows between successive keyframes. The displacements $\Delta x$ and the maximum amplitude $B_{\mathrm{max}}$ are illustrated in~\cref{fig: butterfly_flow_all} (a)(b). By recalculating the WGS with a 0.5-pixel movement, a smoother motion and more stable amplitude can be achieved, as depicted in~\cref{fig: butterfly_flow_all} (c)(d).

\begin{figure}[!ht]
    \centering
    \includegraphics[width=0.5\textwidth]{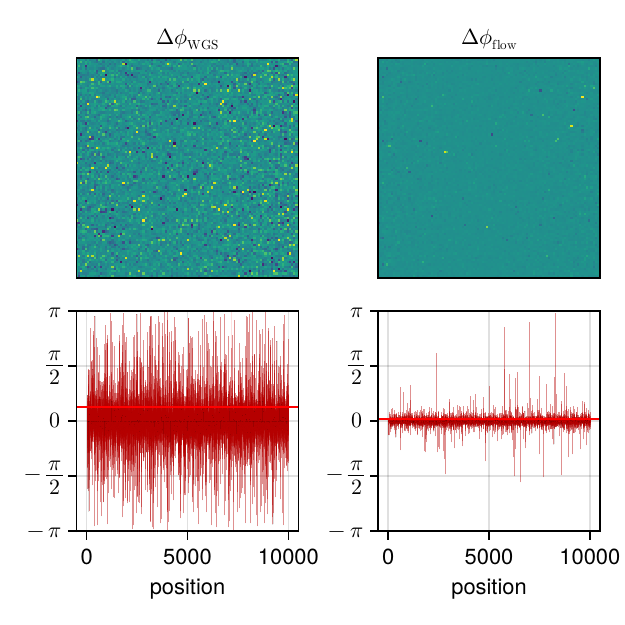}
    \caption{The phase difference $\Delta \phi$ between two consecutive states in the assessment. The lower two histograms represent the unfolding of the phase on the $100 \times 100$ grid shown at the top. The red line denotes the average of the absolute values of $\Delta \phi$. The $\Delta \phi_{\rm{flow}}$ obtained through the current method demonstrates a significantly smoother trend in contrast to $\Delta \phi_{\rm{WGS}}$ computed using the WGS algorithm.}
    \label{fig: butterfly_Delta_phi_wgs_flow}
\end{figure}

Lastly, similar to~\cref{fig: more_points_Delta_phi_wgs_flow}, we evaluate the phase difference between consecutive states in the current analysis. The current gradient flow approach outperforms the pure WGS recalculation method, as depicted in~\cref{fig: butterfly_Delta_phi_wgs_flow}.

\section{Geometric Transformation}
\begin{figure}[H]
    \centering
    \includegraphics[width=0.5\textwidth]{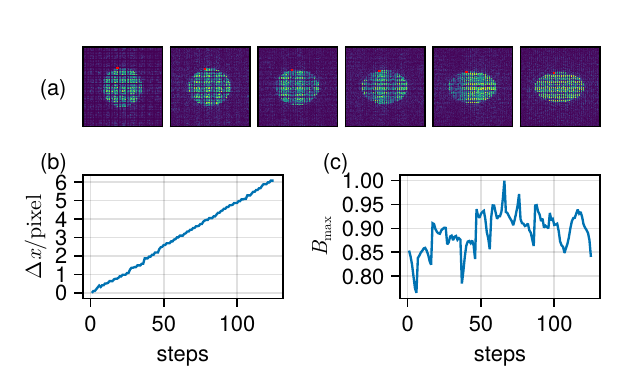}
    \caption{(a) The evolution from a solid circle to an ellipse. (b)(c) The displacement $\Delta x$ per pixel of the central points and the maximum amplitude $B_{\mathrm{max}}$ over successive steps of the red point in (a).}
    \label{fig: circle_to_ellipse_all}
\end{figure}

We arrange a cluster of points into circular pattern and transform it into an elliptical shape. The simulation process is illustrated in~\cref{fig: circle_to_ellipse_all} (a). We find that the image can be adjusted with high efficiency under the gradient flow method even though the contiguous Fourier transformation process is challenging with closely positioned points. The simulation results are shown in~\cref{fig: circle_to_ellipse_all} (b) and (c).

\section{Atoms decay along movement}
\label{sec: atoms_decay}
To investigate heating effects arising from intensity fluctuations during optical trap movement, we developed a one-dimensional semi-classical simulation. The model quantifies the impact of intensity cross-talk that occurs when projecting a trap with a device such as a SLM. The trap's movement is modeled as a discrete step where the potential from the initial trap ($E_{\rm last}$) exponentially decays while the potential of the final trap ($E_{\rm new}$) increases. The total time-dependent potential, $U(t)$, experienced by a single atom is described by the coherent sum of the two electric fields:
\begin{equation}
    U(t) = |E_{\rm last}   {\rm exp}(-t/\tau) + E_{\rm new}   (1 - {\rm exp}(-t/\tau))|^2
\end{equation}
The parameter $\tau$ represents the characteristic transition time and was set to 1 ms to reflect the $ 1~\rm kHz$ frame rate of a state-of-the-art SLM.

\begin{figure*}
    \centering
    \includegraphics[width=0.8\textwidth]{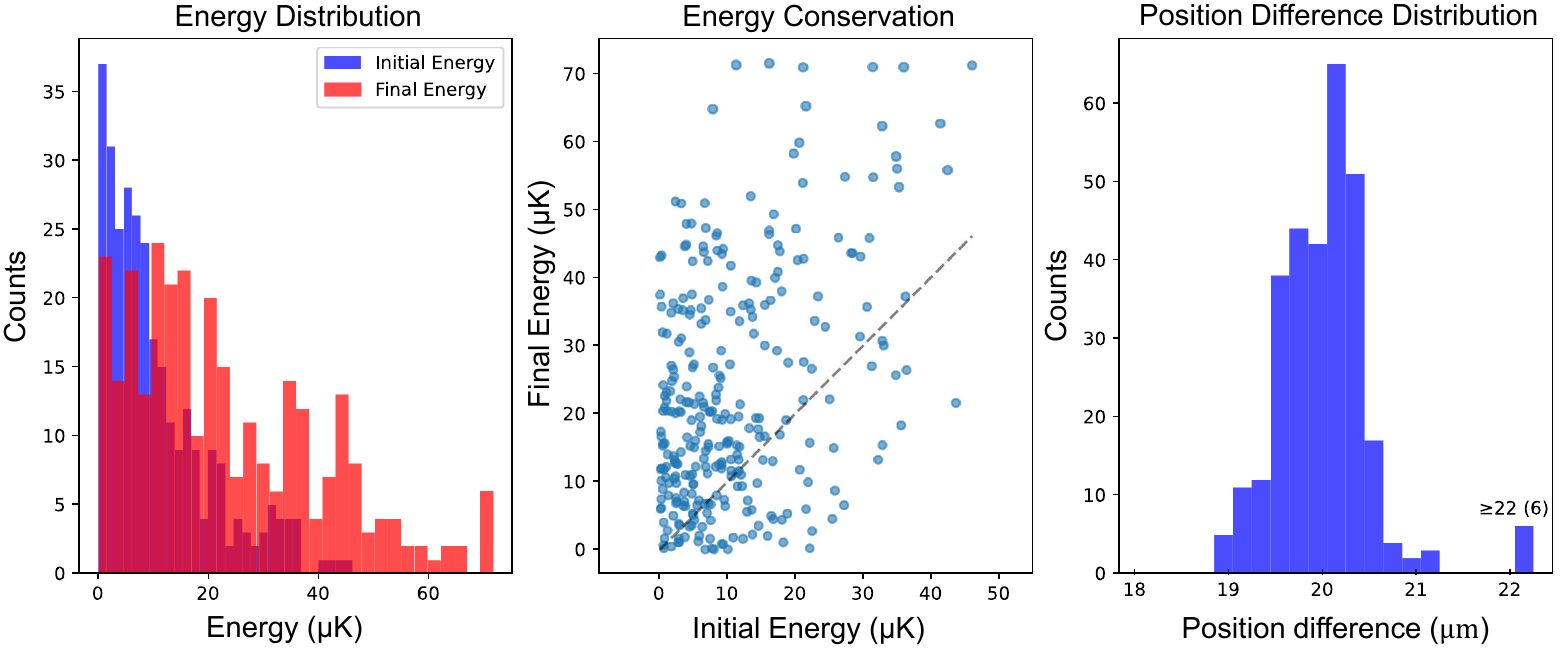}
    \caption{Left: Initial and final energy distributions after displacement. Center: Initial vs. final energy of each atom. Right: Position difference distribution after displacement. Atoms with a final energy $\ge 70~\rm \mu K$ or a position difference $\ge 22~\rm \mu m$ are classified as escaped atoms}
    \label{fig: intensity_fluctuation}
\end{figure*}

The simulation was parameterized for a single Rb atom with an initial temperature of $10~\rm\mu K$, loaded into a trap with a $1~\rm\mu m$ waist and a maximum depth of $0.1~\rm mK$. A Monte Carlo method was employed, where the initial position and velocity of the atom were sampled from a Maxwell-Boltzmann distribution for each of 300 simulation runs. The results, summarized in ~\cref{fig: intensity_fluctuation}, demonstrate a slight increase in the average final energy of the atomic ensemble. Crucially, the calculated atom survival probability remains high, indicating that heating from this translation process is minimal and does not significantly compromise atom confinement.

\newpage
\bibliography{refs.bib}

\end{document}